\documentclass[aps,prl,twocolumn,floatfix,10pt,showpacs]{revtex4-1}

\usepackage{graphicx}
\usepackage{verbatim}
\usepackage{comment}
\usepackage{amssymb}
\usepackage{enumerate}
\usepackage{psfrag}
\usepackage{pstricks}
\usepackage{color}
\usepackage{hyperref}
\usepackage{epstopdf}

\emergencystretch=\maxdimen
\begin{document}

\title{Three-dimensional Hubbard model in the thermodynamic limit}

\author{Ehsan Khatami}
\affiliation{Department of Physics and Astronomy, San Jos\'{e} State University, San Jos\'{e}, CA
95192, USA}

\begin{abstract}
We employ the numerical linked-cluster expansion to study finite-temperature properties of the 
uniform cubic lattice Hubbard model in the thermodynamic limit for a wide range of interaction strengths 
and densities.
We carry out the expansion to the 9th order and find that the convergence of the series extends to lower 
temperatures as the strength of the interaction increases, giving us access to regions of the parameter space 
that are difficult to reach by most other numerical methods. We study the precise trends in the 
specific heat, the double occupancy, and magnetic correlations at temperatures as low as 0.2 of 
the hopping amplitude in the strong-coupling regime. We show that in this regime, accurate estimates for
transition temperatures to the N\'{e}el ordered phase, in agreement with the predicted asymptotic behavior, 
can be deduced from the low-temperature magnetic structure factor. We also find evidence for possible
instability to the magnetically ordered phase away from, but close to, half filling.
Our results have important implications for parametrizing fermionic systems in optical lattice experiments
and for benchmarking other numerical methods.
\end{abstract}


\maketitle

\section{Introduction}

The Hubbard model~\cite{n_mott_49,j_hubbard_63} is the archetypal model for studying 
strongly-correlated systems and their phase transitions in condensed matter physics.
Its Hamiltonian is expressed as
\begin{equation}
H=-t\sum_{\left<ij\right> \sigma}c^{\dagger}_{i\sigma}
c^{\phantom{\dagger}}_{j\sigma} + U\sum_i n_{i\uparrow}n_{i\downarrow}
-\mu\sum_{i\sigma} n_{i\sigma},
\label{eq:H}
\end{equation}
where $c^{\phantom{\dagger}}_{i\sigma}$ ($c^{\dagger}_{i\sigma}$)
annihilates (creates) a fermion with spin $\sigma$ on site $i$,
$n_{i\sigma}=c^{\dagger}_{i\sigma} c^{\phantom{\dagger}}_{i\sigma}$ is
the number operator, $U$ is the onsite Coulomb interaction, $\left<.. \right>$ denotes
nearest neighbors, and $t$ is the corresponding hopping integral. The two-dimensional (2D) 
versions have become {\em de facto} models to describe Mott transition~\cite{m_imada_98}, 
potentially high-temperature superconductivity~\cite{p_anderson_87}, and a wealth of other 
low-temperature phases in materials such as
cuprates~\cite{e_dagotto_94}, pnictides~\cite{g_stewart_11}, iron-selenides~\cite{e_dagotto_13}, 
and heavy fermions~\cite{p_gegenwart_08}, to name a few~\cite{d_scalapino_12}.

Other than a variety of numerical techniques that have been developed over the past three decades 
to solve this often unforgivingly difficult 
model~\cite{j_hirsch_87,r_scalettar_89,k_pan_97,p_kent_05,g_kotliar_06,s_fuchs_11,e_kozik_13},
in recent years, the community has witnessed remarkable efforts in 
simulating, among other correlated theoretical models, the Hubbard model, using ultracold atoms in 
optical lattices~\cite{d_jaksch_98,i_bloch_08}. The observation of the Mott transition with fermions as well as
bosons in different dimensions~\cite{m_greiner_02,t_stoferle_04,i_spielman_07,r_jordens_08,u_schneider_08} 
provided the added impetus for 
attempts to bring down the temperature to a range relevant to superconductivity. 
Even though the latter has proven extremely difficult, there has 
been significant progress along the way.

Some experimental groups have focused on 
the three-dimensional (3D) system~\cite{j_imriska_14,r_hart_15}, where  the N\'{e}el transition to the long-range 
antiferromagnetic (AF) phase is expected to occur at temperatures about an order of 
magnitude larger than that predicted for superconductivity to develop in the two-dimensional 
system~\cite{r_staudt_00,t_maier_05b}. In a groundbreaking study~\cite{r_hart_15}, long-range  AF correlations near the 
critical temperature of the 3D model were observed for the first time. This milestone was reached through
characterization of the experimental system via comparisons of observed thermodynamic properties 
to results from two state-of-the-art numerical techniques, the determinant quantum 
Monte Carlo (DQMC)~\cite{r_scalettar_89} and the numerical linked-cluster expansion (NLCE)
\cite{M_rigol_06,M_rigol_07a,M_rigol_07b,b_tang_13b}. The latter can provide exact results 
in the thermodynamic limit, for a wide range of temperatures and average densities.

The 3D Hubbard model has long been a playground for numerical methods to 
study finite-temperature critical behavior in a system where the electronic correlations can be tuned. 
Its finite-temperature phase transition to the N\'{e}el ordered phase has been studied carefully 
by a variety of numerical 
techniques~\cite{r_scalettar_89,r_staudt_00,p_kent_05,t_paiva_11,e_kozik_13,d_hirschmeier_15}.
However, as these methods are often not well-equipped to handle the strong-coupling regime 
of the model ($U\gtrsim12t$), the complete mapping of the ground state phase diagram 
in the temperature-interaction plane has been assisted by 
an asymptotic behavior based on the critical temperature of the low-energy 
theory in the limit of large interaction strengths~\cite{r_staudt_00,d_hirschmeier_15}. 

Here, we use the NLCE to explore the thermodynamic properties, such as the heat capacity, 
double occupancy, and spin correlations, of the 3D Hubbard model in the thermodynamic limit. 
An outstanding advantage of the NLCE for the Hubbard models over more typical methods is 
that one will have access 
to lower temperatures at larger values of the Coulomb repulsion~\cite{E_khatami_11b}. This, in turn allows us
to explore the divergent behavior of the spin structure factor by using different extrapolation
techniques to obtain reliable estimates for the 
N\'{e}el temperature ($T_N$) in the strong-coupling regime. We find that our estimates match the 
$T_N$ calculated using methods based on quantum Monte Carlo (QMC) within the 
errorbars in the intermediate-coupling regime ($8t\lesssim U\lesssim 12t$),  
and  are in very good agreement with the asymptotic form for $U\gtrsim12t$.
We further confirm our extrapolation schemes in obtaining the N\'{e}el transition temperature for the 3D 
antiferromagnetic Heisenberg model, for which the NLCE can be carried out to significantly higher orders. 
The critical temperature is well known for this model 
 from a large-scale QMC study~\cite{a_sandvik_98}.

\section{the numerical method}

To study Hamiltonian (\ref{eq:H}), we have implemented the NLCE for the 3D cubic lattice.
In the NLCE, an extensive property of the lattice model in the 
thermodynamic limit, normalized to the number of sites, is expressed in terms of
contributions from all finite clusters, up to a certain size, that can be embedded in the lattice. 
These contributions are calculated according to the inclusion-exclusion 
principle. The method can be summarized in the following series:
\begin{eqnarray}
P = \frac{1}{\mathcal{L}} \sum_c W_P(c)
\label{eq:main}
\end{eqnarray}
where $P$ represents the extensive property {\em per site} in the thermodynamic 
limit, $\mathcal{L}$ is the symbolic lattice size ($\to\infty$), and the contribution 
of cluster $c$ to the property, also known as the {\em weight}, is shown by $W_P(c)$. If the 
model does not break  translational symmetry of the underlying lattice, the right-hand 
side of Eq. (\ref{eq:main}) can be simplified to a sum, without the $1/\mathcal{L}$  
factor, over only those clusters that are not related by translational symmetry. 
Further, if the model does not break point group symmetries of the 
underlying lattice, the contribution of all clusters that are 
related by point group symmetry can be expressed as one term that is, the weight of one of the clusters,
times a {\em multiplicity} factor, which represents the number of ways one can obtain a distinct 
cluster by applying point group symmetry operations to that cluster.

Equation (\ref{eq:main}) is a cluster expansion~\cite{m_sykes_66} that can be written 
not only for the infinite lattice, but also for a finite cluster~\cite{b_tang_13b}. 
We use this fact to find the weights. Consider, for example, the equation for $p(c)$, 
the property calculated for a finite cluster $c$:
\begin{eqnarray}
\label{eq:inclusion}
p(c) = W_P(c) + \sum_{s\subset c} W_P(s),
\end{eqnarray}
where we have intentionally separated the weight of $c$ itself, with  $s$ running over
all subclusters of $c$ (clusters obtained by removing different number of sites from $c$). Note that 
$p(c)$ is not normalized by the number of sites. By reordering the 
terms in this equation, we can write the weight of each cluster as its property less the 
contributions of its subclusters:
\begin{eqnarray}
\label{eq:exclusion}
W_P(c) = p(c) - \sum_{s\subset c} W_P(s).
\end{eqnarray}
The above equation provides a recursive method for calculating all the weights
up to a certain size; We start with the smallest cluster, a single site, which does not
have any subclusters, i.e., $W_p(1)=p(1)$, and generate larger clusters  
by adding sites to it one by one in the so-called {\em site expansion} NLCE. We 
carry out this expansion to the 9th order, which means we will work with clusters 
as large as 9 sites.

NLCEs use the same basis as the
high-temperature series expansions (HTSEs)~\cite{linked}. However, the calculation of the
extensive quantities at the level of individual clusters [$p(c)$] is left to an
exact numerical method, such as the exact diagonalization, as opposed to a
perturbative expansion in terms of inverse temperature, as done in the HTSEs. 
Despite the lack of an explicit small parameter, having a finite 
series inevitably leads to the loss of convergence below a certain
temperature, where the correlations in the system extend beyond a length
of the order of the largest sizes considered.  However, the exact
treatment of clusters leads to minimum convergence temperatures that are
lower than those of HTSE with a comparable  order (see Fig.~\ref{fig:Cv}). 

We study the specific heat, $C_v$, which can be obtained within
NLCE from the knowledge of the energy, density, and their correlation, 
without any numerical derivatives or numerical integration~\cite{e_khatami_12b},
double occupancy, $D=\left<n_{\uparrow}n_{\downarrow} \right>$ (site averaged), 
the nearest-neighbor spin correlations,
\begin{equation}
S^{zz} = \frac{1}{\mathcal{L}} \sum_{\left<ij \right>} \left< S^z_i S^z_j \right>,
\end{equation}
where $S^z_i$ is the $z$ component of the spin operator at site $i$, and 
the spin structure factor,
\begin{equation}
S({\bf q}) = \frac{1}{\mathcal{L}} \sum_{jk} e^{i{\bf q}\cdot ({\bf r}_j-{\bf r}_k)}\left< S^z_j S^z_k \right>,
\end{equation}
at ${\bf q} = (\pi,\pi,\pi)$, where ${\bf r}_j$ is the displacement vector for site $j$ on the lattice. 
We denote the latter by $S^{AF}$. In the disordered phase at high temperatures, it remains 
finite in the thermodynamic limit, while its divergence at low temperatures
is the indication for AF ordering in the system. 

Similar to the analytic Pad\'{e} approximations used extensively in
HTSEs, here we take advantage of two {\it numerical} resummation
techniques to improve the convergence of our series at low temperatures.
We use the Euler algorithm~\cite{Euler} to resum the last 6 terms of
the series or the Wynn algorithm~\cite{Wynn} with 3 and 4 cycles of
improvement (details about these techniques and their use for NLCEs can be found in
Refs.~\onlinecite{M_rigol_07a,b_tang_13b}). Except for the staggered structure factor,
for which we know the Euler algorithm does not perform as well as the Wynn 
(see the discussion surrounding Fig.~\ref{fig:SAFT}), we take the average of properties from
the last two orders after the Euler, and the last orders of each of the Wynn
resummations as our best estimate. To quantify our confidence in
the accuracy of the resummed results, we define a ``confidence region''
around this average where all four values that contribute to the average
fall.  Thus, the errorbars in our figures simply mark the boundaries of
this region and should not be mistaken as statistical errorbars.

\section{Results}

\begin{figure}[t]
\centerline {\includegraphics*[width=3.3in]{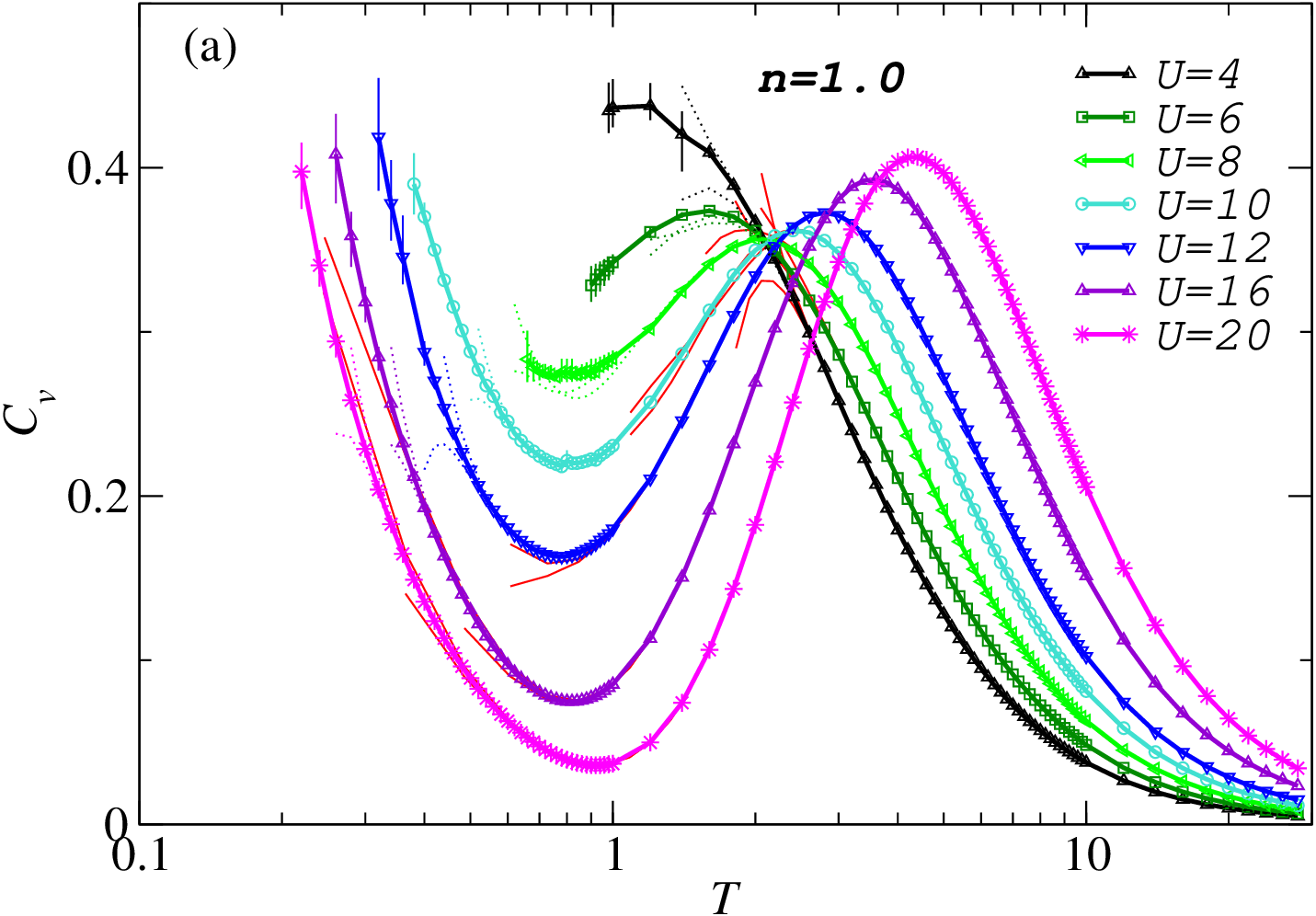}}
\centerline {\includegraphics*[width=3.3in]{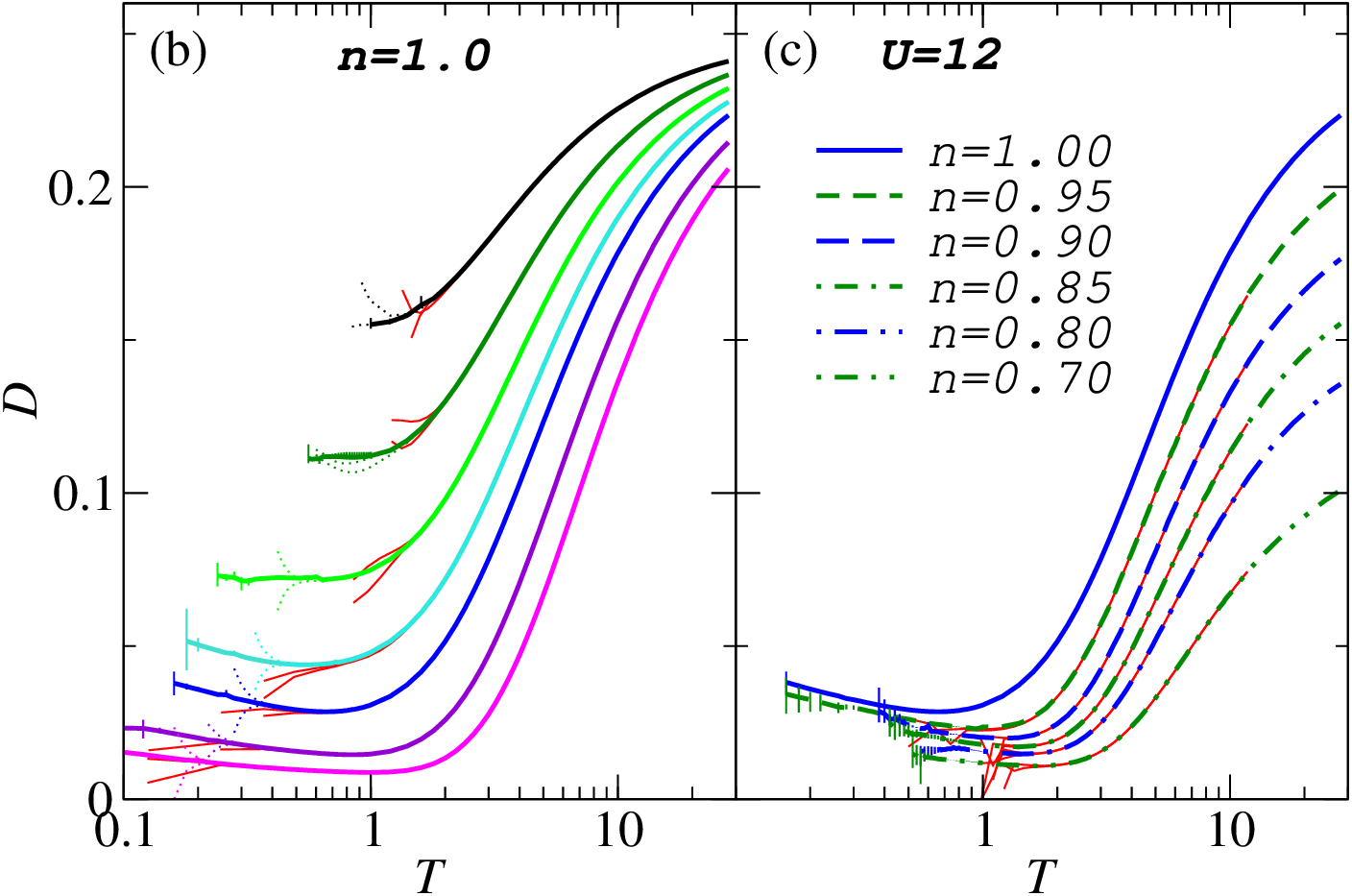}}
\caption{(a) Specific heat of the 3D Hubbard model at half filling vs temperature for several values 
of the interaction strength. Thick solid lines represent the average between the last two orders after Euler
resummation of the last 6 terms of the series, and Wynn resummations with 3 and 4 cycles of improvement.
The errorbars mark the confidence region where 
all the values used in the average fall. Thin dotted lines are the last two orders of the bare NLCE
sums, and thin solid lines are the 9th and 10th orders of the HTSE~\cite{v_scarola_09,r_jordens_10,l_deleo_11} 
(b) Double occupancy at half filling vs temperature. 
$U$ increases from top to bottom with the same values as in (a). 
(c) Double occupancy at $U=12$ away from half filling vs temperature.}
\label{fig:Cv}
\end{figure}

The NLCE makes the study of thermodynamic quantities in the thermodynamic limit efficient and easy. In Fig.~\ref{fig:Cv}, 
we show the specific heat of the half-filled system as a function of temperature for an interaction 
strength that ranges from $U=4$ in the weak-coupling regime to $U=20$ in the strong-coupling 
regime. We have set $k_{\rm B}=1$, and $t=1$ as the unit of energy throughout the paper.
Since the properties of the clusters in the series are calculated using exact diagonalization,
we obtain information at all temperatures and chemical potential values for a given $U$ in a 
single run, as opposed to QMC-based methods in which each temperature and chemical
potential has to be treated separately. For this reason, one 
can choose a fine temperature and chemical potential grid to better capture the details and trends in 
quantities of interest in different regions of the parameter space (see Fig.~\ref{fig:Cv}). This is also of great importance for 
modeling of systems in optical lattice experiments. For instance, because of the existence 
of a trapping potential, theoretical modeling of the inhomogeneous system is often
done through proper averaging of properties over a set of homogeneous systems whose chemical 
potentials vary only slightly from one to the next. 
This is known as the local density approximation. The other advantage of exact diagonalization
is that one has full access to the partition function of the clusters, and so, there is no need to employ
numerical integration or derivations, which can introduce systematic errors, for the calculation of $C_v$, 
or the entropy.

As can be seen in Fig.~\ref{fig:Cv}, we capture the exact location of the high-temperature peak 
in $C_v$ for all values of $U$ shown, and reach temperatures as low as $0.2$ for the largest $U$ of 20. 
The high-temperature peak marks the temperature region where local moments form as the system is cooled.
Like for the 2D counterpart~\cite{e_khatami_12b}, and as expected from the increasingly dominant Mott
physics, this region moves to higher temperatures as the interaction is increased. On the other hand, 
the minimum convergence temperature of the series decreases
as $U$ increases. In the strong-coupling regime, this temperature is proportional to the exchange interaction 
of the effective Heisenberg model, which scales as $1/U$. One can see that the lowest convergence
temperatures achieved with the NLCE before resummations (thin dotted lines in Fig.~\ref{fig:Cv}) are generally 
lower than those achieved using the HTSE up to the 10th order (red thin solid lines), especially in the intermediate-coupling regime, $U=8-12$. 

A feature of $C_v$ that can be resolved here with high accuracy is the unique crossing
point of curves for different $U$ around $T=2$, which persists for $U\lesssim 12$. 
The physical implications of this phenomenon, which is
ubiquitous in strongly-correlated systems and has also been observed to persist
for $U$ up to the bandwidth in the two-dimensional version of the Hubbard
model~\cite{t_paiva_01,e_khatami_12b} are discussed in Ref.~\onlinecite{d_vollhardt_97}.

\begin{figure}[t]
\centerline {\includegraphics*[width=2.8in]{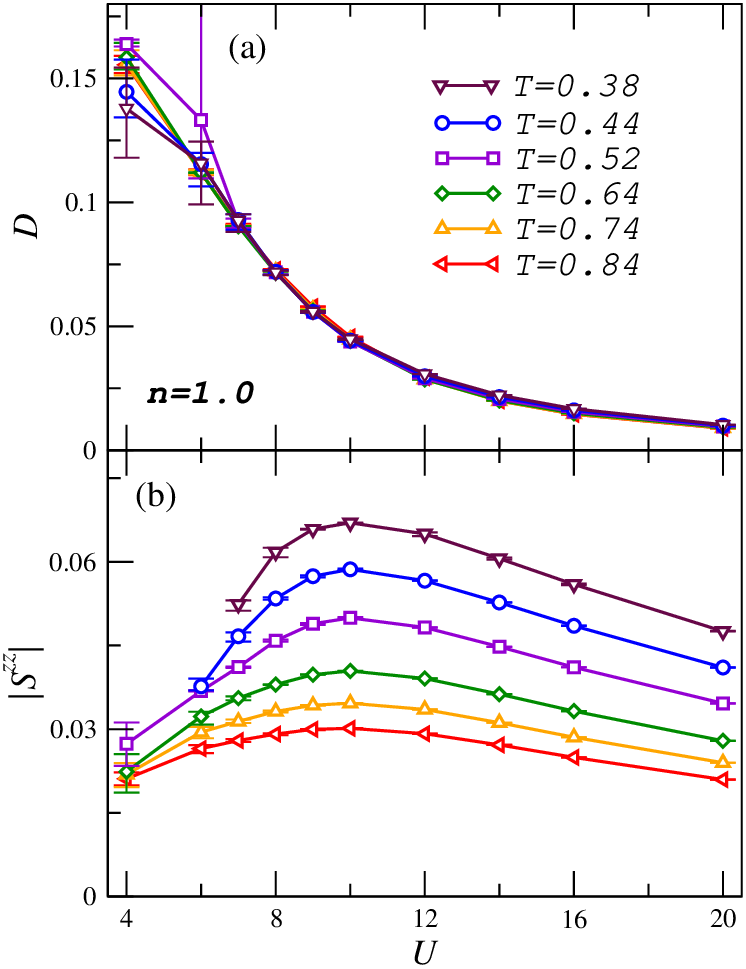}}
\caption{(a) Double occupancy at half filling as a function of the interaction strength at different 
temperatures. $D$ rapidly decreases by increasing the interaction strength as one approaches 
the Mott phase. (b) Nearest-neighbor spin correlations at half filling vs $U$. Similarly	 to the 
2D model~\cite{E_khatami_11b}, they initially rise in the weak-coupling regime by increasing 
the interaction strength and eventually 
decrease as $1/U$ in the strong-coupling regime. Its peak 
around $U=10$ becomes more pronounced as the temperature is lowered.}
\label{fig:D}
\end{figure}

In Figs.~\ref{fig:Cv}(b) and \ref{fig:Cv}(c), we show the double occupancy at, and away from, half filling
as a function of temperature. At half filling, the uncorrelated limit of $D$ at high temperatures is $\left<n_{i\uparrow} 
\right>\left<n_{i\downarrow} \right>=\frac{1}{4}$ regardless of $U$. However, the larger the $U$, the faster 
$D$ drops upon decreasing the temperature. Quantum fluctuations leave the system with a nonzero
and $U$-dependent double occupancy even at $T=0$. Away from half filling, the uncorrelated values of $D$
vary as $\frac{n^2}{4}$ with the electron density, $n$, and the ground state values are expected to remain 
nonzero. A clear upturn in $D$ upon decreasing the temperature is found after the initial drop at all 
fillings, at least with $U\geq 10$ at the accessible temperatures. 
Close to half filling, this phenomenon can be attributed to the tendency of the system to order 
antiferromagnetically, thus, giving way to a larger number of allowed virtual hoppings to 
neighboring sites that were otherwise forbidden by the Pauli's exclusion principle in the uncorrelated 
system~\cite{e_gorelik_10,E_khatami_11b}.
Sufficiently far from half filling, or in the weak-coupling regime~\cite{t_paiva_11}, it is believed that 
the upturn is associated with Fermi liquid physics.\cite{f_werner_05,k_mikelsons_09,E_khatami_11b} 
Our results match those obtained using DQMC with clusters as large as $8^3$~\cite{t_paiva_11}; however, 
in the weak-coupling regime, we are limited to $T\gtrsim 1$.

\begin{figure}[t]
\centerline {\includegraphics*[width=3.3in]{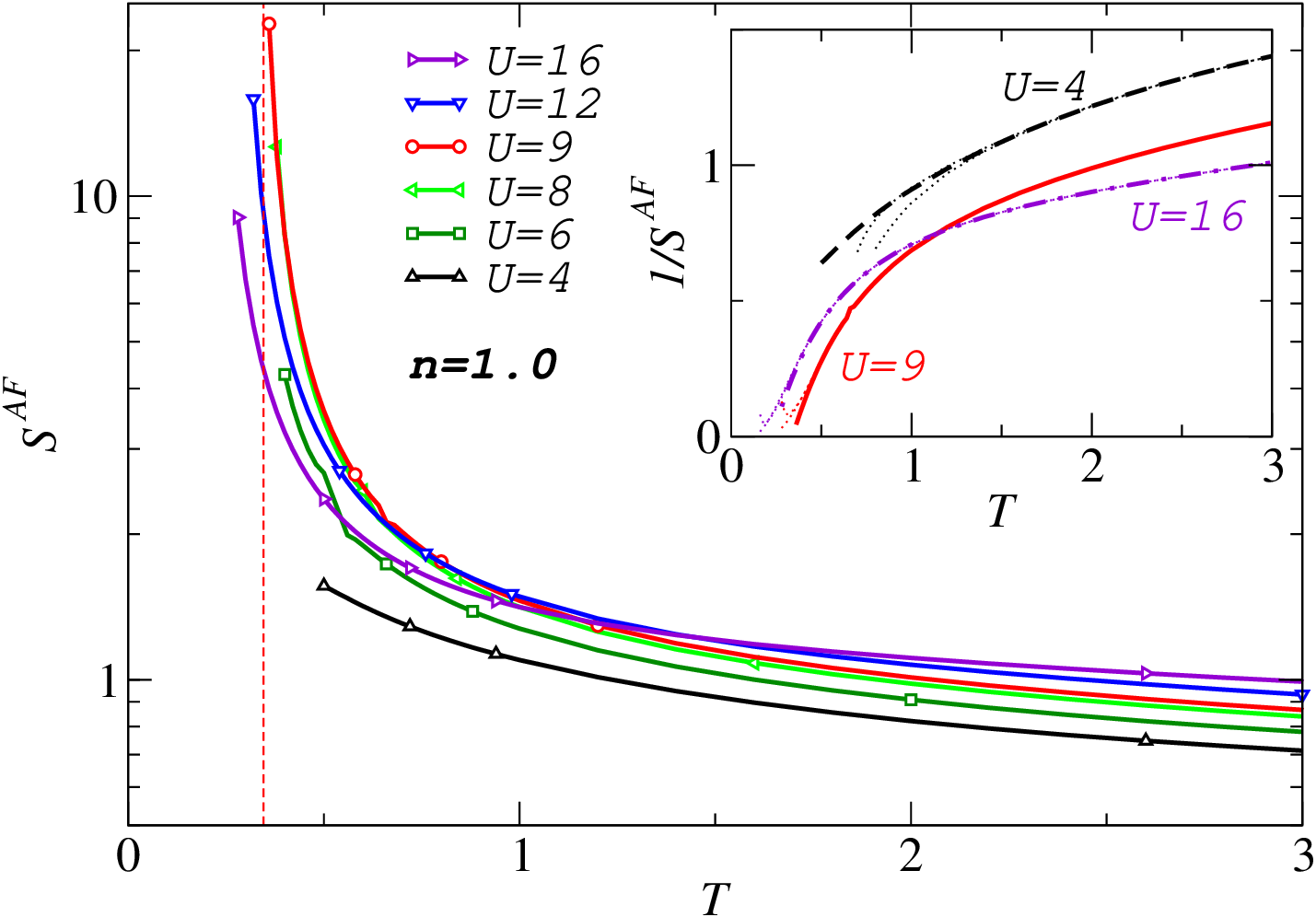}}
\caption{Antiferromagnetic structure factor at half filling as a function of temperature for several 
values of the interaction strength. Wynn resummation with 3 and 4 cycles of improvement have 
been used. The results are shown for temperatures above where the two estimates match within 
a few percent.  The low-temperature structure 
factor increases as $U$ increases to 9, then reduces upon further increasing of the strength 
of the interaction. The dashed vertical line indicates the location of the transition temperature for $U=9$.
Inset shows the inverse of $S^{AF}$ for select values of $U$. A simple fit
to $A/(T-T_N)^B$ for $T<0.6$ suggests $T_N=0.35$ for $U=9$, which is consistent with current 
best estimates. Thin dotted lines show bare NLCE results for the last two orders.}
\label{fig:SAFT}
\end{figure}

The change in the double occupancy as we vary $U$ can be better seen in Fig.~\ref{fig:D}(a), 
where we show $D$ at half filling as a function of $U$ in a low-temperature window. The large
fluctuations of data at $U<8$ point to the lack of convergence in the series at the lowest 
temperatures. Regardless, the double occupancy clearly decreases rapidly by increasing $U$ with 
no sharp features or outstanding variation in its behavior as the temperature is decreased from $T=0.84$ 
to $T=0.44$. In the same time, short-range AF correlations display a non-monotonic 
behavior. The absolute value of $S^{zz}$, shown in Fig.~\ref{fig:D}(b), initially increases with 
increasing $U$, and then slowly decreases as $U$ further increases in the strong-coupling 
regime. This behavior can be understood by considering the interplay between moment formation,
which takes place at higher temperatures for larger values of $U$, and the strength of the 
exchange interaction between moments, which decreases with increasing $U$. 
The overall behavior is reminiscent of that for the 2D system~\cite{E_khatami_11b}.  We find that
the peak at $U=10$ becomes sharper as the temperature is decreased, hinting that short-range 
correlations may eventually be strongest at a slightly larger interaction strength than the one corresponding 
to the largest N\'{e}el transition temperature to the long-range order ($U=9$)
~\cite{r_staudt_00,p_kent_05,t_paiva_11,g_rohringer_11,e_kozik_13,d_hirschmeier_15}.

\begin{figure}[t]
\centerline {\includegraphics*[width=3.3in]{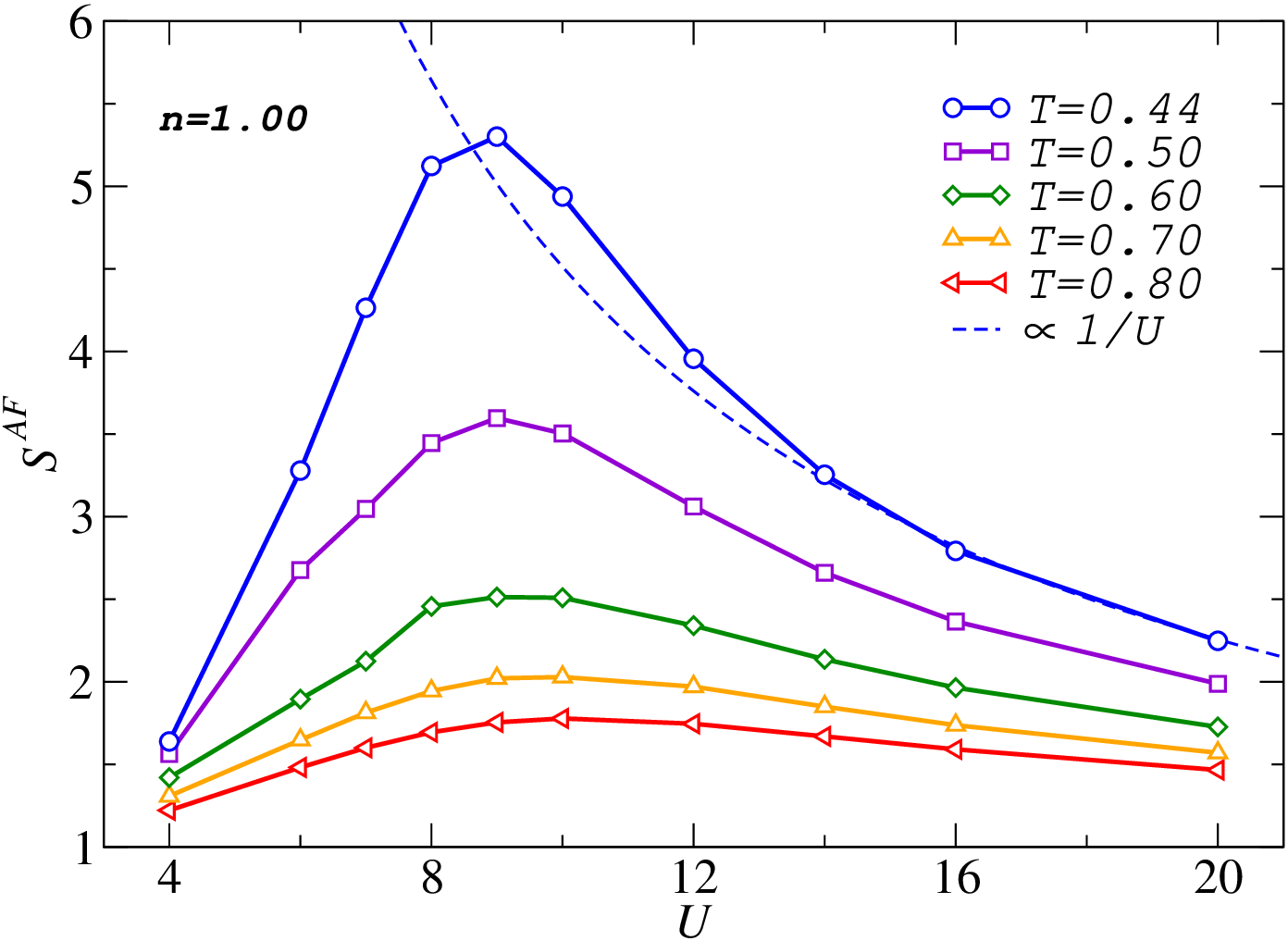}}
\caption{Antiferromagnetic structure factor at half filling as a function of the interaction strength 
at different temperatures. The results are shown after Wynn resummations with 4 cycles of 
improvement (they also match results from Wynn with 3 cycles of improvement roughly within the 
size of symbols). The dashed line is a fit to $A/T$ (constant $A$) using values of 
$S^{AF}$ for the three largest $U$'s at $T=0.44$, representing the theoretical asymptotic trend 
in the strong-coupling regime.}
\label{fig:SAFU}
\end{figure}

In Figs.~\ref{fig:SAFT} and \ref{fig:SAFU} we explore the AF structure factor, 
$S^{AF}$, which contains information about correlations
at all length scales. In Fig.~\ref{fig:SAFT}, we show the temperature dependence of $S^{AF}$
for various interaction strengths. The Euler resummation technique behaves poorly for fast 
growing properties such as $S^{AF}$ whose terms in the series at a given $T$ do not necessarily 
alternate in sign. For this reason, we perform only 
Wynn resummations with 3 and 4 cycles of improvement for the structure factor and show the 
latter at temperatures where the two are within roughly 10\% of each other. The reliability of 
the Wynn resummations for $S^{AF}$ have been confirmed also through previous comparisons to DQMC 
results~\cite{r_hart_15}. It is worth noting that the lowest convergence temperature decreases 
by increasing $U$, something we already saw for $C_v$ in Fig.~\ref{fig:Cv}. Of course, this 
does not coincide with a larger $S^{AF}$ at low temperatures for a larger $U$ in the strong-coupling regime. In fact, 
the extent of the correlations in the system, which essentially controls the convergence of the NLCE,
is becoming smaller as the interaction is getting stronger.

The model has a finite-temperature phase transition to the 
long-range N\'{e}el ordered state. Therefore, the structure factor for any $U\ne 0$ is expected 
to diverge at a nonzero temperature. The transition temperature, $T_N$, is expected to be largest around $U=9$.
So, it is not surprising to find that $S^{AF}$ is also largest for $U=9$ at the lowest accessible temperatures. 
In the strong-coupling regime, $T_N$ is expected to be around $J$ ($\sim 4t^2/U$)
\cite{a_sandvik_98}, the strength of the exchange interaction in the effective low-energy 
Heisenberg model. The inset of Fig.~\ref{fig:SAFT} shows the inverse of $S^{AF}$ for $U=4$, 9, 
and 16. A simple fit to $A/(T-T_N)^B$ (constant $A$ and $B$) for $T<0.6$ suggests that $T_N\sim0.35$ 
for $U=9$, which is consistent with current best estimates~\cite{t_paiva_11,g_rohringer_11,e_kozik_13,d_hirschmeier_15}.
It is also evident that $T_N$ will be
smaller for the other two values of $U$ in the inset. The critical temperature as a function of $U$ 
and its different estimates within the NLCE will be discussed below.

Similarly to the nearest-neighbor correlations, the structure factor as a function of $U$, plotted in
Fig. ~\ref{fig:SAFU}, exhibits a peak at $U=9$, which develops faster than that for the former
 as the temperature is lowered. This is an indication of the fast growing long-range correlations 
 in the system as one approaches the critical 
temperature. The dashed line in Fig.~\ref{fig:SAFU} is a fit proportional to $J$ ($\propto 1/U$) using the structure 
factor for the largest three $U$ values at $T=0.44$. It makes clear the asymptotic behavior of the 
magnetic correlations in the strong-coupling regime.

\begin{figure}[t]
\centerline {\includegraphics*[width=3.3in]{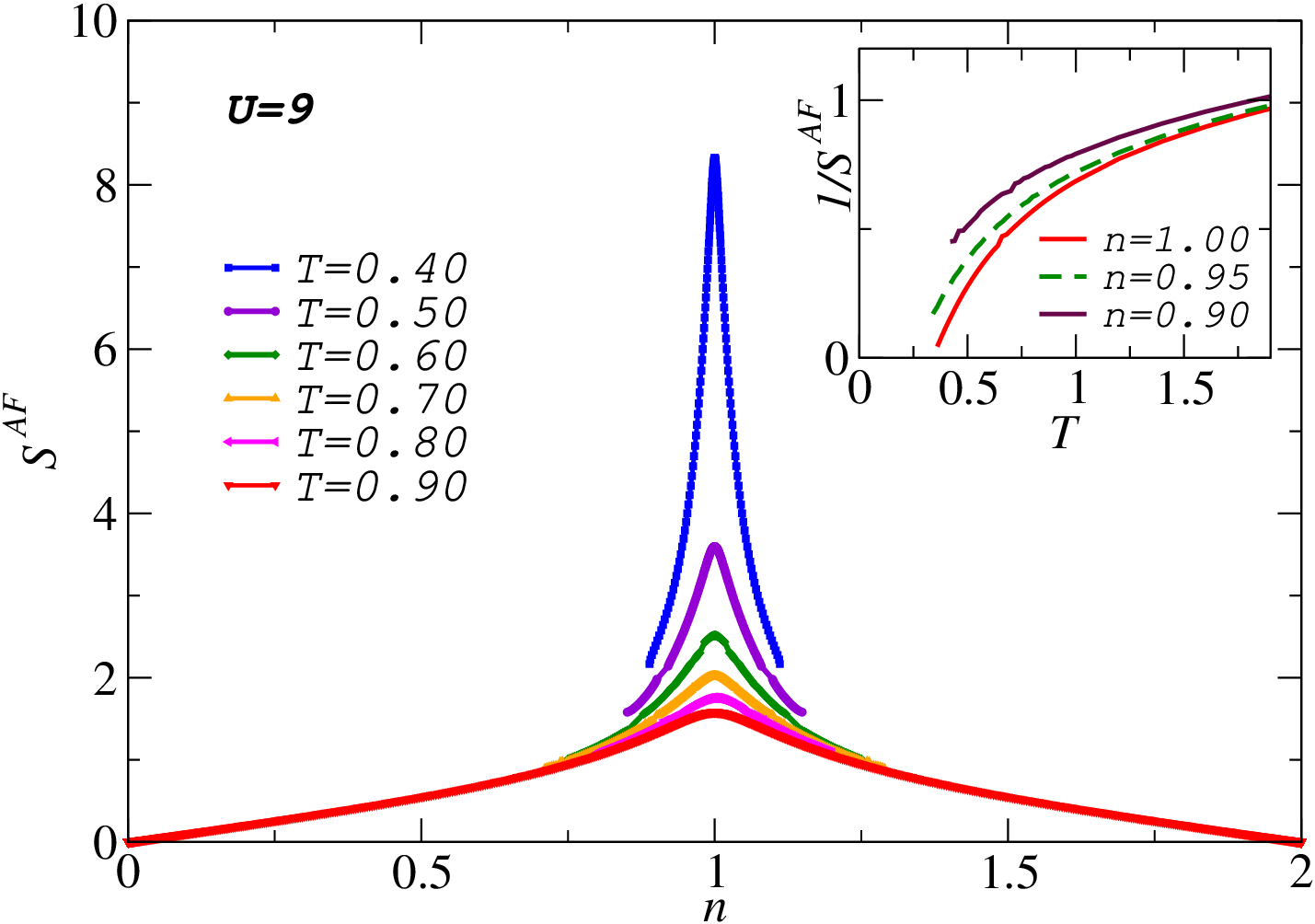}}
\caption{Antiferromagnetic structure factor with $U=9$ as a function of density at several 
temperatures. As the temperature is lowered, the structure factor develops a sharp peak around 
half filling. Inset: Inverse of the structure factor vs temperature at three different densities.}
\label{fig:SAFn}
\end{figure}

So far, we showed results for the structure factor only at half filling. 
But, what happens to the divergent AF correlations in a system with a $n\ne 1$? 
To answer this question, we plot in Fig.~\ref{fig:SAFn} $S^{AF}$
as a function of density for $U=9$ at different temperatures. A very sharp peak develops at
$n=1.00$ as the temperature is lowered, indicating that the correlations in the system remain 
large only in the close proximity of half filling. These results are of special importance for the 
simulation of the model using ultracold fermionic
atoms in optical lattices as a range of densities are present simultaneously at different radii from
the center of the trap~\cite{r_hart_15}. 

The system can in principle make a transition to the 
long-range N\'{e}el phase even away from half filling. To explore this possibility, 
we plot $1/S^{AF}$ as a function of temperature for different $n$ in the inset of Fig.~\ref{fig:SAFn}. 
The structure factor at $n=0.95$ shows strong indication of a nonzero critical temperature that is
nonetheless, smaller than that for the half-filled system. At a smaller density of $n=0.90$,
we do not have enough low-temperature data from the NLCE to draw any conclusion about 
the critical temperature. We point out that despite the divergent behavior of $S^{AF}$ close to half 
filing, an instability to a different type of order may be dominant in this region.
We have not studied such a scenario here.

\begin{figure}[t]
\centerline {\includegraphics*[width=3.3in]{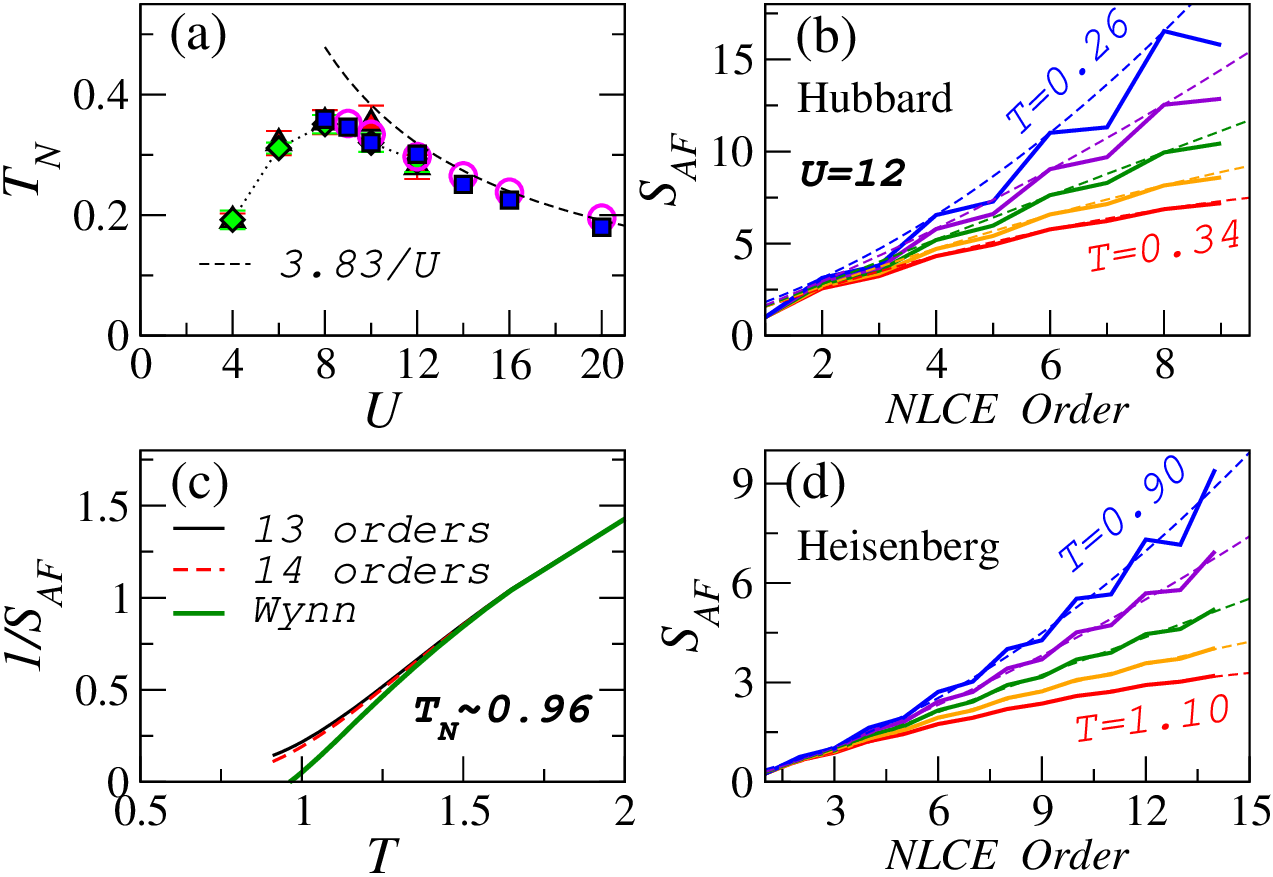}}
\caption{(a) Estimated N\'{e}el transition temperature vs the interaction strength. Filled squares
are obtained from extrapolations of the NLCE $S^{AF}$ to low temperatures, and empty circles are obtained
from fits to $S^{AF}$ as a function of the NLCE order (see text). Filled triangles and diamonds are data from 
the DCA~\cite{p_kent_05} and DQMC~\cite{r_staudt_00}, respectively. The dotted line is a guide to the eye.
The dashed line is the theoretical asymptotic function.
(b) The bare AF structure factor for $U=12$ vs the NLCE order
at different temperatures (with a uniform grid) above and below  the expected transition temperature ($T_N\sim0.30$).   
Dashed lines represent fits of results in the even orders to a 2nd-degree polynomial. 
(c) Inverse of $S^{AF}$ as a function of temperature for the 3D Heisenberg model with $J=1$ [which 
sets the unit of energy in (c) and (d)].
The last two orders of the bare sums and results after Wynn resummations with 6 cycles of 
improvement are shown. The latter points to a divergence at $T_N\sim0.96$. 
(d) The staggered structure factor for the Heisenberg model as a function
of the NLCE order at different temperatures (in a uniform grid) around $T_N$. 
Here, dashed lines are
2nd-degree polynomial fits of data in all orders.}
\label{fig:TN}
\end{figure}

As we saw in Fig.~\ref{fig:SAFT}, the critical temperatures can be estimated from the 
extrapolations of the structure factor in the intermediate- to strong-coupling regime,
where enough information at low temperatures are available. In Fig.~\ref{fig:TN}(a), we
plot the N\'{e}el temperatures deduced in this way as a function of $U$ as filled circles.
We also plot $T_N$ for $U\le 12$ from the dynamical cluster approximation (DCA)~\cite{m_hettler_98,p_kent_05} 
and DQMC, which match our results within the errorbars. The NLCE results are in very 
good agreement with the theoretical prediction for the large-$U$ Heisenberg 
limit for $U> 12$ as well~\cite{i_affleck_88,a_sandvik_98}.

In the ordered phase, we expect the maximum $S^{AF}$ to
scale linearly with the cluster size, $N$, for finite clusters since the correlations 
extend to all sites. On the other hand, in the disordered phase above $T_N$, $S^{AF}$
increases with $N$ linearly so long as the linear size of the cluster is smaller in order than
the correlation length. For larger systems, $S^{AF}$ as
a function of $N$ saturates to a temperature-dependent value. Within our NLCE, the 
order refers to the size of the largest clusters in the expansion. Thus, the 
order of the expansion does not exactly represent $N$ as in a finite-size calculation 
due to the existence of a large number of smaller clusters in the series. However, 
the role of the latter is to eliminate boundary effects~\cite{s_sahoo_15}, and so, 
one could expect that the expansion order would approximately play the role of $N$. 
In fact, the NLCE order has been successfully used as a length scale to study the scaling 
of R\'{e}yni entropies at quantum critical points~\cite{a_kallin_13,a_kallin_14,e_stoudenmire_14,s_sahoo_15,t_devakul_15}.

With this assumption, we plot in Fig.~\ref{fig:TN}(b) the bare sums
(partial sums without resummations) of $S^{AF}$ for $U=12$ vs 
the NLCE order at five different temperatures, ranging from $T=0.26$ to $T=0.34$. 
The fluctuations from one order to the other are smaller at higher $T$, where the convergence 
is achieved faster. For $T>T_N$, we expect the function to be 
weaker than linear, and eventually saturate to a finite value in the thermodynamic 
limit. On the other hand, for $T<T_N$, it is expected to show at least a linear overall increase 
by increasing the order and to diverge in the thermodynamic limit. In other words, the critical
temperature can be located by monitoring the  overall curvature of $S^{AF}$ as a 
function of NLCE order as it changes sign from negative above $T_N$ to positive below
$T_N$. Hence, we fit the data to a second-degree polynomial and locate $T_N$ as the 
temperature where the quadratic coefficient of the polynomial vanishes. 
For $U\le 10$, the partial sums fluctuate wildly between even and odd orders at temperatures
near the expected $T_N$, and the fits largely underestimate $T_N$ as compared to 
available estimates. At $U=10$, we obtain a $T_N$ that is about 10\% less 
than the average of DCA and DQMC estimates.  We find that if instead we perform the 
fits to data only using even orders [shown in Fig.~\ref{fig:TN}(b) as dashed lines], the 
agreement between the resulting $T_N$ and those from the extrapolations of the 
structure factor to low temperatures, and the DCA and DQMC estimates, improves in 
the intermediate-coupling region. The former are plotted in Fig.~\ref{fig:TN}(a) 
as open circles. Our results are within the errorbars of the results of the
DQMC and the DCA, as shown in that figure as triangles and diamonds.
The difference between $T_N$'s obtained from fits to all orders and fits to only 
even orders of the NLCE decreases as the interaction increases in the strong-coupling region.
For $U=16$, this difference is only about 3\%.

Neglecting odd orders in the fits is an arbitrary choice, which is likely due to the fact that we have 
a small number of terms in the series. Obviously, we cannot obtain results at higher orders for the Hubbard model. 
However, we are mostly focused on the strong-coupling regime of this model whose approximate 
low-energy theory is the Heisenberg model~\cite{i_affleck_88}. The Heisenberg
model has the advantage of having a much smaller Hilbert space, which 
makes it possible for us to go to significantly higher orders in the NLCE. We
have carried out the NLCE for the 3D Heisenberg model up to the 14th order.
The resulting inverse $S^{AF}$ is plotted in Fig.~\ref{fig:TN}(c), where both the last two 
orders of the bare sums and the Wynn resummation after 6 cycles
of improvement can be seen. The latter matches with the resummation
after 5 cycles all the way to the transition temperature for this model, which is
found to be $0.96J$. This value is close to that obtained by finite-size scalings in 
a QMC study ($T_N=0.95J$)~\cite{a_sandvik_98}. Now, the question is: can we 
arrive at a similar $T_N$ with fits to the $S^{AF}$ vs the NLCE order? We find 
that  here, with more terms in the expansion, regardless of whether we fit the 
polynomial using even or all orders of the NLCE, we find 
$T_N=1.01$. This value is about 6\% more than that obtained from QMC. However, this is 
also roughly the error between the different fitting schemes and between the NLCE and DCA/DQMC estimates
for $T_N$ in the Hubbard model when $U\sim 12$. 

In summary, we have implemented a NLCE in three dimensions, up to the 9th order in the site expansion, 
to study the exact thermodynamic
and critical behavior of the Hubbard model in the thermodynamic limit. We study trends in the 
specific heat, double occupancy and magnetic correlations in the model as we tune the 
strength of the interaction. We find that both short-range and long-range AF correlations 
are largest around $U=9-10$ at the lowest temperatures available. We further extract 
the N\'{e}el temperature by extrapolating the AF structure factor to lower temperatures and 
find strong evidence that the instability to the AF phase persists at densities close to half filling.
We also explore a different scheme in which polynomial fits to bare partial sums of the 
series for the structure factor can provide accurate estimates of the transition temperature in the strong-coupling 
region of the model. We confirm this method by extracting $T_N$ for the 3D Heisenberg model, where
we can obtain a larger number of terms. 
This scheme can be exploited in future to study critical phenomena in other models.

We thank Vito W. Scarola for providing the high-temperature series expansion results and are grateful to 
Rajiv.~R.~P.~Singh, Richard T. Scalettar and Marcos Rigol for their insightful comments on the manuscript.
This work used the Extreme Science and
Engineering Discovery Environment (XSEDE) under Project No.
TG-DMR130143, which is supported by NSF Grant No. ACI-1053575.
This work was partly supported by Research Scholarship and Creative Activity
grants at San Jos\'{e} State University.

\end{document}